\documentclass[twocolumn,showpacs,amssymb,aps,nofootinbib,floatfix]{revtex4}
\usepackage{epsfig}
\newcommand{\be}{\begin{eqnarray}}
\newcommand{\ee}{\end{eqnarray}}
\newcommand{\ave}[1]{\left\langle #1 \right\rangle}

\begin{document} \hbadness=10000
\topmargin -0.8cm\oddsidemargin = -0.7cm\evensidemargin = -0.7cm
\title{Scaling of $v_2$ in heavy ion collisions}
\author{Giorgio Torrieri}
\affiliation{Institut f\"ur Theoretische Physik,
  J.W. Goethe Universit\"at, Frankfurt A.M., Germany  
torrieri@fias.uni-frankfurt.de}
\date{June 10, 2007}  

\begin{abstract}
We interpret the scaling of the corrected elliptic flow parameter
w.r.t. the corrected multiplicity, observed to hold in heavy ion collisions
for a wide variety of energies and system sizes.
We use dimensional analysis and power-counting arguments to place
constraints on the changes of initial conditions in systems with
different center of mass energy $\sqrt{s}$.   Specifically, we show
that a large class of changes in
the (initial) equation of state, mean free path, and longitudinal geometry
over the observed $\sqrt{s}$ are likely
to spoil the scaling in $v_2$ observed experimentally.
We therefore argue that the system produced at most Super Proton
Synchrotron (SPS) and Relativistic Heavy Ion Collider (RHIC)
energies is fundamentally the same as far as the soft and approximately
thermalized degrees of freedom are considered.  The ``sQGP'' (Strongly
interacting Quark-Gluon Plasma) phase, if it is there,
is therefore not exclusive to RHIC.  We suggest, as a goal for further
low-energy heavy ion experiments, to search for a ``transition''
$\sqrt{s}$ where the observed scaling breaks.
\end{abstract}
\maketitle
%%%%%%%%%%%%%%%%%%%%%%%%%%%%%%%
\section{Introduction}
The azimuthal anisotropy of mean particle momentum (parametrized by it's
second Fourier component $v_2$), thought of as originating from the
azimuthal anisotropy in collective flow (``elliptic flow''), has long been regarded as an
important observable in heavy ion collisions.
The main reasons for this is that elliptic flow has long been understood
to be ``self-quenching'' \cite{v2orig,v2orig2}: The azimuthal pressure gradient
extinguishes itself soon after the start of the hydrodynamic evolution, so the final
$v_2$ is insensitive to later stages of the fireball evolution and
therefore allows us to probe the hottest, best thermalized, and
possibly deconfined phase.

In addition, as has been shown in \cite{teaney}, the $v_2$ signature
is highly sensitive to viscosity.  The presence of even a small but
non-negligible viscosity, therefore, can in principle be detected by a
careful analysis of $v_2$ data.

Indeed, one of the most widely cited (in both the academic and popular
press) news coming out of the heavy ion community concerns the discovery, at RHIC, of a ``perfect fluid'',
also sometimes referred to as ``sQGP'' 
\cite{whitebrahms,whitephobos,whitestar,whitephenix}.
The evidence for this claim comes from the successful
modeling of RHIC $v_2$ by boost-invariant hydrodynamics
\cite{heinz,shuryak,huovinen}.   The scaling of $v_2$ according to the number of constituent quarks further suggests that the flow we are seeing is partonic, rather than hadronic, in origin \cite{npart1,npart2}, especially since the scaling applied to kinetic energy (rather than transverse momentum) holds for every known species up to the lowest momentum \cite{phenv21,starv21,taranenko}.

\begin{figure}[h]
%\begin{center}
%\epsfig{width=6cm,clip=1,angle=-90,figure=fig_v2phobos.ps}
%\epsfig{width=6cm,clip=1, angle=-90,figure=systemtable.eps}
%\hspace{1cm}
\epsfig{width=10cm,clip=1,figure=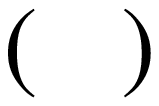}
\hspace{-2cm}
\caption{(color online)\label{uni}A compendium of evidence for the universal (for
  the observed energies) scaling of $v_2/\epsilon$ vs
  $\frac{1}{S}\frac{dN}{dy}$ \cite{starexp}.  Predictions from ideal
  boost-invariant hydro are also shown (Lines, see references of \cite{starexp}).}
%\end{center}
\end{figure}

While hydrodynamics is a fully deterministic theory, it contains a crucial not very
well understood assumption: initial conditions.   While the
degree of boost invariance is not currently well known experimentally,
the transverse structure of the energy density should follow a
Glauber model \cite{glauber}, based on the superposition of the initial 
nuclear
densities.
That allows us to characterize the collision in terms of a number less
than one called the eccentricity $\epsilon$, related to the impact
parameter $b$ and the radius R ($\sim A^{1/3}$fm)
\begin{equation}
\label{epsilondef}
\epsilon = \frac{\sqrt{2R+b}-\sqrt{2R-b}}{\sqrt{2R+b}}
\end{equation}
the total transverse area of the system $S$ depends, similarly, on $R$ and $b$
\begin{equation}
\label{sdef}
S = 2 R^2 \cos^{-1} \left( \frac{b}{2 R} \right) - b\sqrt{R^2 - \frac{b^2}{4}}
\end{equation}
Since the announcement of the discovery of the perfect fluid, a considerable amount of high quality
experimental data has been collected.  In particular, extension of
RHIC beams to smaller colliding systems such as $Cu-Cu$ have allowed us
to compare systems of similar multiplicity but in very different
energy regimes.  The results have been remarkable: It seems that
$v_2/\epsilon$ 
(where $\epsilon$ is the initial eccentricity), plotted
against $\frac{dN}{S dy}$ (where $S$ is the area of the collision
system and $\frac{dN}{dy}$ is the multiplicity rapidity density),
fall on a ``universal'' curve, which links very different regimes,
ranging from Alternating Gradient Synchrotron (AGS) to RHIC
(\cite{phobosexp,starexp},Fig. \ref{uni}).  

This scaling, albeit with a linear dependence rather than the slightly curved one observed, has been predicted previously \cite{voloshin1,voloshin2}
on the basis of a nearly ``free streaming'' calculation where the mean free path is comparable to the system size.   Such a limit considerably under-predicts the observed $v_2/\epsilon$, which is why a nearly-perfect hydrodynamic regime is thought to apply.   As we will show, however, the same scaling in this regime is far from guaranteed.
%%%%%%%%%%%%%%%%%%%%%%%%%%%%%%
%%%%%%%%%%%%%%%%%%%%%%%%%%%%%%%

We do not possess at present the
tools,such as 3D viscous hydrodynamics, to quantitatively analyze this data.
However, the extent of the scaling suggests that these very different
systems vary somehow only in one scale, and that this scale is
connected to the total entropy produced \cite{caines}.
%%%%%%%%%%%%%%%%%%%%%%%%%%%%%%
\begin{figure}[h]
\begin{center}
%\hspace{-1cm}
%\epsfig{width=6cm,clip=1, angle=-90,figure=systemtable.eps}
%\hspace{1cm}
\epsfig{width=9cm,clip=1,figure=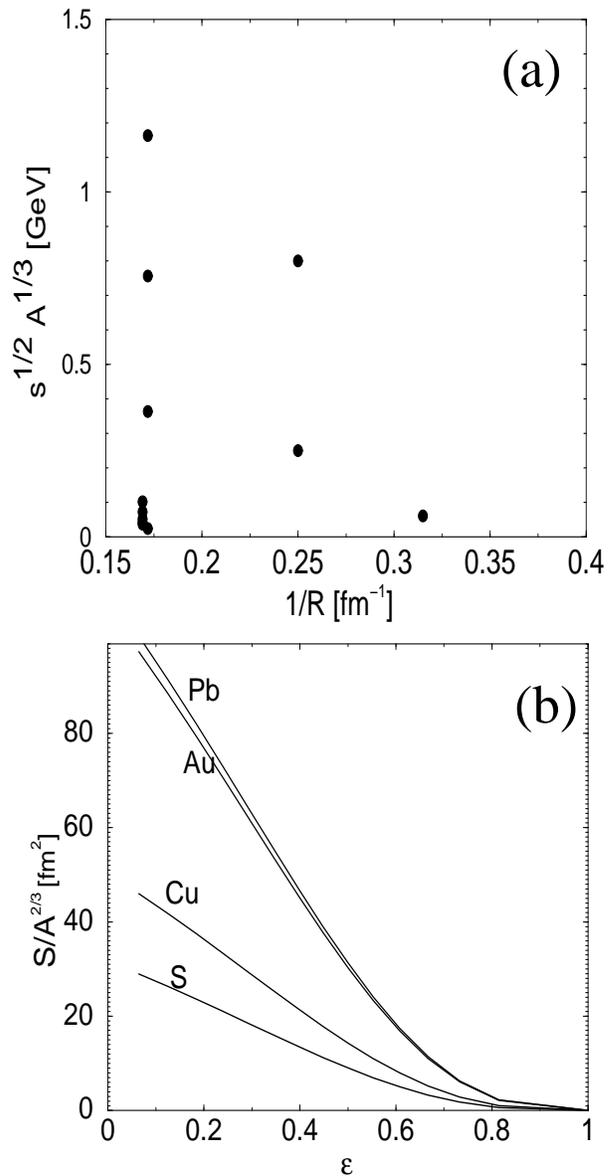}
\caption{(color online)\label{range}
Top panel (a):  The range in $\sqrt{s} A^{1/3}$ vs $1/A^{1/3}$ for the
  available experimental data  Bottom panel (b): The range of $S$ and $\epsilon$
  covered by available data}
\end{center}
\end{figure}
%%%%%%%%%%%%%%%%%%%%%%%%%%%%%%%

In this work, we shall perform a \textit{ qualitative} analysis, using
 elementary tools such as Taylor expansion, dimensional analysis,
and power counting.   We show that these tools, together with
experimental data, allow us to place stringent limits on the initial
conditions of the system
created in heavy ion collisions, from AGS to RHIC energies.

In particular, we show that the observed scaling places very strong
constraints on initial microscopic
properties (entropy density, mean free path), as well as longitudinal
structure.  We argue that statistical and transport properties  can
not significantly vary between RHIC and observed lower energies.  
 We therefore
conclude that the ``perfect fluid'', if it is there, is a common
characteristic of all experimentally studied systems so far.
%------------------

It should be underlined that the ``experimental result'' on which we base our conclusions is, itself, somewhat theory-laden since $\epsilon$ is not an experimentally measured quantity.   If the best physical description of the soft initial dynamics at RHIC is not the Glauber model but rather, for example, the color glass condensate \cite{cgcorig}, the scaling might need to be revisited \cite{cgc}.
Even within the Glauber model calculation, it is only when eccentricity fluctuations \cite{eccfluct} are taken into account that the universal scaling is observed (one of the reasons why this scaling was not reported until recently, and missed for example in \cite{v2130}).  
Yet, the fact that an improvement in the computation of $\epsilon$ results in improvement of the scaling is indicative that something physical must be behind it. 

It should be noted that the result in Fig. \ref{range} is based on $p_T$ integrated
$v_2$, and is therefore subject to systematic uncertainties due to the wide range of
acceptance in the experiments summarized in Fig. \ref{range}. Recent investigations
(Fig. 1 of \cite{scalingphenix,scalingptphobos}), however, seems to show that the 
scaling is {\em not}
dependent on integrating over $p_T$, since a compatible scaling holds when separate
$p_T$ bins are considered.  The universal scaling was found to hold in rapidity space
also \cite{scalingptphobos}
  
In the rest of the paper, we will regard the result as established because we wish to show that its consequences are profound.  Finding simple scaling in a system as complicated as a heavy ion collisions is encouraging enough that it deserves exploration even through the scaling observation should be regarded as preliminary.  We hope that the experimental community will soon determine how universal this simple scaling really is.

\section{How natural is the observed scaling within hydrodynamics?}
For the system under consideration, three dimensionless parameters can safely be
assumed to be significantly smaller than unity: The initial spatial
eccentricity $\epsilon$, the initial speed of sound $c_s$ (that
encodes information on the equation of state \cite{flork,florkcs,friman}), and the
initial mean free path $l_{mfp}$ divided by the initial size of
the system,in general parametrized by the transverse radius $R$ and the
longitudinal size $\ave{z}$.  The latter refers to the system's
longitudinal size in configuration space, at mid-rapidity, \textit{at the start} of the hydrodynamic evolution,
and is a general definition. It is usually thought that $\ave{z}$ can interpolate from a Landau initial
condition \cite{landau}, where $\ave{z}$ is related to the nucleon
mass $m_N$ and the center of mass energy $\sqrt{s}$ by 
\begin{equation}
\ave{z} \sim R\frac{\sqrt{s}}{m_N}
\end{equation}
, to the
Bjorken \cite{bjorken} initial condition, where, for a
system with maximum rapidity $y_L$, 
\begin{eqnarray}
 \ave{z} \sim 2 \tau_0 \sinh (y_L) \\
y_L = \frac{1}{2} \ln \left( \frac{\sqrt{s}+\sqrt{s-m_N^2}}{\sqrt{s}-\sqrt{s-m_N^2}} \right) 
\end{eqnarray}
 where $\tau_0$
is the thermalization timescale of the system. 

Since hydrodynamic evolution is fully determined by the
initial conditions and the equation of state\footnote{ An additional
variable experimental observables can depend on is a \textit{ freeze-out
  criterion}, encoded by, for example, $\frac{T_{freeze-out}}{T_{initial}}$.
We disregard this variable in the subsequent discussion as the
observables we discuss should not strongly depend on it.}
,  any flow variable
can be thought of as a function of the parameters characterizing
these.
If this function is integrable (i.e., if no turbulence occurs), than
it is safe to expand this function around any dimensionless parameter
less than one.

Both the dimensionless $v_2$ and the dimensionful
$\frac{dN}{S dy}$ ($\sim fm^{-2}$) can therefore be Taylor-expanded
around these quantities.  For $v_2$, we know that the 0th term is 0
(perfectly central collisions have no $v_2$), so
\begin{eqnarray}
v_2 \sim a_{100} \epsilon + a_{200} \epsilon^2 + \nonumber \\
      \epsilon \left( a_{110}^R \frac{l_{mfp}}{R} +a_{110}^z
      \frac{l_{mfp}}{\ave{z}}+  \right)  \nonumber +\\
      \epsilon c_s \left( a_{111}^R \frac{l_{mfp}}{R} +a_{111}^z
      \frac{l_{mfp}}{\ave{z}}
      \right) +  ...
\label{eqv2}
\end{eqnarray}
$a_{i j k}$ are in general (probably transcendental) functions of 
an arbitrary number of dimensionless quantities constructed out 
of $\ave{z}$, $T$,$\mu_B$, $R$ (in general $\zeta = \sum_{mnl} \zeta_{mnl}$, where
$\zeta_{mnl}=\ave{z}^{m} T^{n} \mu_B^{l} R^{l+n-m}$, all $m,n,l$). 
The exact form of $a_{ijk}$ can be obtained  integrating the hydrodynamic equation from the initial
time to freeze-out time (through they are expected to be insensitive to the
latter, and equivalently to  $\frac{T_{freeze-out}}{T_{initial}}$).
Ideal boost-invariant hydrodynamics, with a bag model or
lattice-inspired equation of state \cite{shuryak,heinz,huovinen},
predicts that $a_{20j} \ll a_{10j}$, and hence $v_2/\epsilon$ is
approximately constant (as the lines in Fig. \ref{uni} show). 
 
The experimentally observed rise of $v_2/\epsilon$ with
multiplicity, and encounter with the hydrodynamic calculation (see
Fig. \ref{uni}), can
therefore be interpreted as RHIC energy being the only point where the
system reaches the ``ideal hydrodynamics limit''.

It is however unclear to what extent is such a conclusion an artifact
of the models being used to perform this comparison assuming
exact \cite{huovinen,shuryak,heinz} or approximate
\cite{hirano,kodama} boost-invariance as an initial condition at all
energies.  While there are good physical arguments for why such an
initial condition is appropriate for heavy ion collisions
at mid-rapidity \cite{bjorken}, the fact that some experimental data is
more compatible with Landau hydrodynamics even at RHIC highest
energies \cite{steinberg} suggests the need to question this
assumption, and in particular to evaluate it's effect on our estimate
of the transport properties \textit{and their energy dependence}.
While no viscous calculation using Landau initial conditions has so
far appeared in the literature, it is reasonable to suppose that the
slower cooling from a Landau initial condition would leave more time
for $v_2$ to form.  Hence, the limits on viscosity/mean free
path/thermalization time inferred from $v_2$
data are strongly correlated with the degree of assumed boost invariance.

$\frac{dN}{S dy}$ contains information about both the longitudinal
structure at freeze-out and the  \textit{final} particle number
density (a function of the initial $T$ and $\mu_B$).
In an ideal (isenthropic) expansion, the final entropy is equal to the
initial entropy content of the system ($\sim$ the initial particle
density $n(T,\mu_B)$), so
\begin{equation}
\frac{1}{S}\frac{dN}{dy} \sim \ave{z} n(T,\mu_B)
\end{equation}
Collective evolution, if the system has a non-negligible mean free path, can however create additional entropy. 

The first correction to isenthropic expansion should therefore be
proportional to the entropy creation due to viscous processes.  This
is given by \cite{landaubook}
\begin{equation}
 \Delta S \sim \int dt \eta \ave{\partial_\mu u_{\nu}}^2 \frac{V(t)}{T}
\end{equation}
where $\eta$ is the viscosity, $u_{\nu}$ the flow field, and $V(t)$
the volume of the fluid.

 viscosity is in turn proportional to the mean free path,
density $n$, and mean momentum current $\ave{p}$ \cite{landau}
\begin{equation}
\eta \sim l_{mfp} n(T,\mu_B) \ave{p}(T,\mu_B)
\end{equation}
since the initial volume is, to 0th order in eccentricity, $\sim \ave{z} R^2$,we get
\begin{eqnarray}
\frac{1}{S} \frac{dN}{dy} \sim \ave{z} n(T,\mu_B) \left( 1 +  b_{010}^R \frac{ l_{mfp}}{R} + b_{010}^{\ave{z}} \frac{ l_{mfp}}{\ave{z}}
 + \right. \nonumber \\
\epsilon \left( b_{110}^R \frac{ l_{mfp}}{R} + b_{110}^{\ave{z}} \frac{ l_{mfp}}{\ave{z}}
   \right)+ \nonumber \\ \left.
c_s \left( b_{011}^R \frac{ l_{mfp}}{R} + b_{011}^{\ave{z}} \frac{
 l_{mfp}}{\ave{z}} \right) + ... \right)
\label{eqn}
\end{eqnarray}
where, once again, the coefficients $b$ are functions of any
arbitrary sum $\sum_{m,n,l} \ave{z}^{m} T^{n} \mu_B^{l} R^{n+l-m}$, that have to be
calculated by integrating the viscous hydrodynamic equations.
Since entropy is predominantly produced in
the initial collisions\cite{decoherence}, we can again disregard  $\frac{T_{freeze-out}}{T_{initial}}$.

It is immediately clear that several of the parameters used in the
previous expansions (in particular the $c_s,l_{mfp},\ave{z}$) can not, by
causality, depend on the \textit{transverse} system size $A$, and have to depend
only on the local energy density \textit{only}.  If soft
observables scale with the number of participants, therefore, these
parameter's dependence on energy and system size can only be a
function of $\sigma$, where
\begin{equation}
\sigma = f(\sqrt{s}) A^{1/3} \sim \sqrt{s} A^{1/3}
\end{equation}
   Not all parameters, however, have this dependence: The
initial transverse system size $R$ and eccentricity
$\epsilon$ exhibit no energy dependence.  Rapditiy \cite{scalingptphobos} provides a further indipendent direction.

Presently available heavy ion experiments have explored a significant
range in $\epsilon,\sigma$ and $R$, shown in Fig. \ref{range}.
Had $v_2/\epsilon$ scaled in a different way from  $\frac{1}{S}\frac{dN}{dy}$ 
w.r.t. these variables, the currently available experimental data
should have signaled it by the appearance of ``branches'', systems
with same $\frac{1}{S}\frac{dN}{dy}$ but different $v_2/\epsilon$ (or vice-versa).
Hence, the absence of such branches is a probe
capable of constraining the nature of the system
created in heavy ion collisions.   In the rest of the paper, we will
use this probe, as well as Eqs \ref{eqv2} and \ref{eqn}, to constrain the system as much as possible.

\begin{figure}[h]
\begin{center}
\epsfig{width=8cm,clip=1,figure=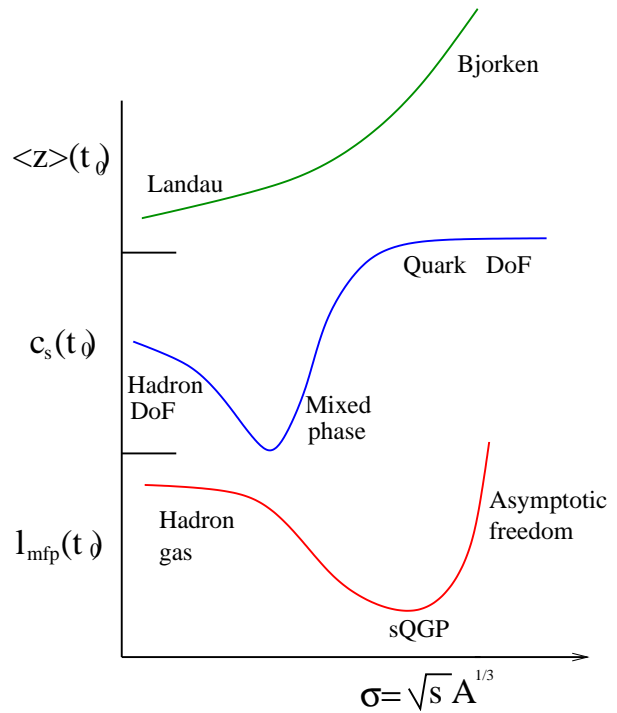}
\caption{(color online)\label{sqrts}The expected dependence, with energy of the
  initial $\ave{z},c_s$ and $l_{mfp}$.  These quantities,by
  definition, will be independent of system size}
\end{center}
\end{figure}

Fig. \ref{sqrts} sketches the dependence of $c_s,l_{mfp}$ and $\ave{z}$ as
conventionally thought.  As the energy increases, $c_s$ rapidly dips
in the ``mixed phase''\footnote{In a first order phase transition, it
dips to zero.  In a cross-over, it merely decreases \cite{florkcs}}
and than increases as the system moves from a Hadron Gas to a Quark
Gluon Plasma. An increase in $c_s$, if other variables were fixed, 
would \textit{increase} the $v_2/\epsilon$ (since there is more
pressure build-up) and, since it's
an equilibrium process, either maintain $\frac{dN}{S dy}$ constant or
decrease it (if longitudinal flow is significant).
%%%%%%%%%%%%%%%%%%%%%%%%%%%%%%
%%%%%%%%%%%%%%%%%%%%%%%%%%%%%%%

$l_{mfp}$ is sensitive to the phase structure in a somewhat different
manner: the system should first enter the strongly coupled low
viscosity sQGP regime, and
then, when $\sigma$ is very high ($\sqrt{s}\sim$ TeV), the higher viscosity
high-Temperature regime where the QGP becomes asymptotically free.
A higher viscosity means a lower $v_2$ \cite{teaney}, but a higher
$\frac{dN}{S dy}$, as more entropy is created in the system and
collective energy is transformed, by microscopic interactions, into
thermal energy.

Finally, the stopping power of the system should decrease as the
initial condition goes from Landau \cite{landau} to the Bjorken \cite{bjorken} limit.  This
decreases the pressure build-up
needed to create $v_2$ as well as $\frac{1}{S} \frac{dN}{dy}$, since
entropy is re-distributed in a wider rapidity
space.
%%%%%%%%%%%%%%%%%%%%%%%%%%%%%%
\begin{figure}[h]
\begin{center}
\epsfig{width=9cm,clip=1,figure=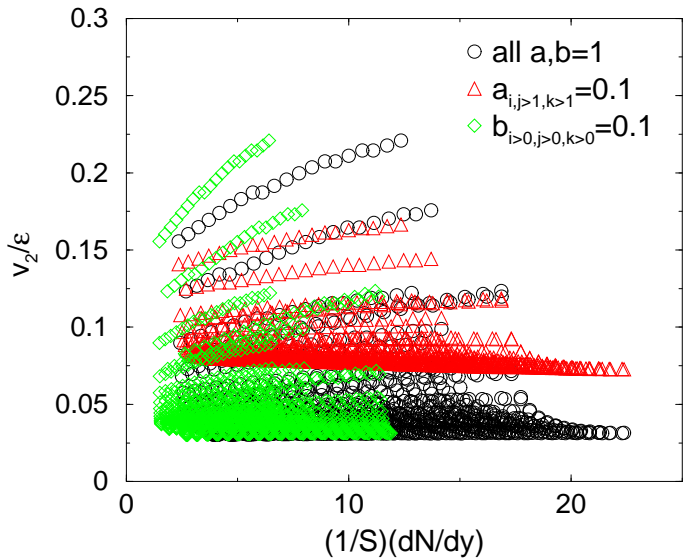}
\caption{(color online)\label{scalefile} Scaling expected with for ``typical''
  values of $a_{ijk},b_{ijk}$
  and the formula in \cite{milov} for multiplicity.  The bands, lower
  to upper, correspond to different classes in centrality (10) and
  different system sizes (Pb,Au,Cu,S). }
\end{center}
\end{figure}
%%%%%%%%%%%%%%%%%%%%%%%%%%%%%%%
The combined effect of these three scalings is difficult to
evaluate without a solver of 3D viscous hydrodynamic equations.  It is however likely to be
non-negligible, and depend \textit{ only} on $\sigma$, \textit{ not}
on A or centrality/$\epsilon$.   The total size of the system, and it's lifetime, however,
should depends strongly on \textit{both} $\sigma$ and $A$, as well as
the eccentricity.

Hence, terms $\sim \frac{l_{mfp}}{R,\epsilon R,\ave{z}}$, present in
both Eq. \ref{eqv2} and \ref{eqn} should vary
in a non-trivial way as energy and system size are changed.  In addition, for
systems with a short lifetime, or a slow formation of $v_2$, terms
$\sim \epsilon^{i>1}$ in Eq. \ref{eqv2} should become non-negligible.
Thus, the
appearance of a common scaling between systems with different
$\sigma,A,\epsilon$,such as the one observed appears highly unlikely \textit{a priori}.

We shall further explore the naturalness of the observed scaling using
\textit{natural values}
of $a_{ijk},b_{ijk}$, together with parametrizations of $c_s$ and
$l_{mfp}$
\begin{eqnarray}
\label{eostoy}
\rho = \frac{\rho_{HG}}{2} \left( 1+ \tanh(T_c -T)  \right) +\nonumber \\ \frac{\rho_{QGP}}{2
  } \left( 1 + \tanh(T- T_c) \right) 
\end{eqnarray}
\begin{eqnarray}
\label{cstoy}
c_s = \frac{0.1}{2} \left( 1+0.9 \tanh(T_c -T)  \right) +\nonumber \\ \frac{c_s^{ideal}}{2
  } \left( 1 + \tanh(T- T_c) \right) 
\end{eqnarray}
\begin{eqnarray}
\label{lmfptoy}
\frac{l_{mfp}}{fm} = \frac{5}{2} \left( 1+\tanh(T_c -T)  \right) +\\ \nonumber \frac{0.1}{2
  } \left( 1 + \tanh(T-T_c)\right)*\log \left( 1+ \frac{T}{T_c} \right) 
\end{eqnarray}
We then use the Bjorken formula to
get the initial longitudinal and energy density distribution
\begin{equation}
\frac{1}{S}\frac{dN}{dy} \sim \frac{1}{\pi A^2} \frac{dN}{dy} = \tau_0 \rho(T,\mu_B)
\end{equation}
where $\frac{dN}{dy}$ can be obtained from experiment by a
phenomenological formula \cite{milov}.
\begin{equation}
\frac{dN}{dy} = \frac{N_{Participants}}{1.48} \ln \left( \frac{\sqrt{s}}{1.48\mathrm{GeV}} \right)
\end{equation}
The temperature is found by solving the resulting conservation of
energy equations, using the equation of state given in Eq. \ref{eostoy}.
The resulting temperature is then plugged into Eqs. \ref{lmfptoy} and
\ref{cstoy} to calculate $c_s$ and $l_{mfp}$.  These, together with
``typical'' coefficients $a_{ijk},b_{ijk}$ and $R \sim
A^{1/3},\ave{z}\sim \tau_0 \sinh (y_L)$
 are then used to
investigate how $v_2/\epsilon$ and $\frac{1}{S}\frac{dN}{dy}$ depend
on each other.

Since all ``small'' dimensionless parameters are encoded in
$c_s,\epsilon,\frac{l_{mfp}}{R,\ave{z}}$, all coefficients
$a_{ijk} \sim \left. v_2 \right|_{exp},b_{ijk} \sim 1$.
Fig. \ref{scalefile} shows what kind of scaling is to be expected from
Eqs \ref{eqv2} and \ref{eqn} if 
the approximate equalities hold exactly.  The branch structure is
clearly seen across the experimentally studied system sizes.
It is of course possible to eliminate and produce universal scaling,
but, in the absence of a deeper principle why that should be so, the
coefficients $a,b$ would need to be carefully \textit{fine-tuned}.

Thus, the experimentally observed scaling of Fig. \ref{uni} places
very profound constraints on how the microscopic properties, and the
global longitudinal structure, can vary between AGS and RHIC energies.
\section{What drives the scaling?}
\begin{figure*}[h]
\begin{center}
\epsfig{width=19cm,clip=1,figure=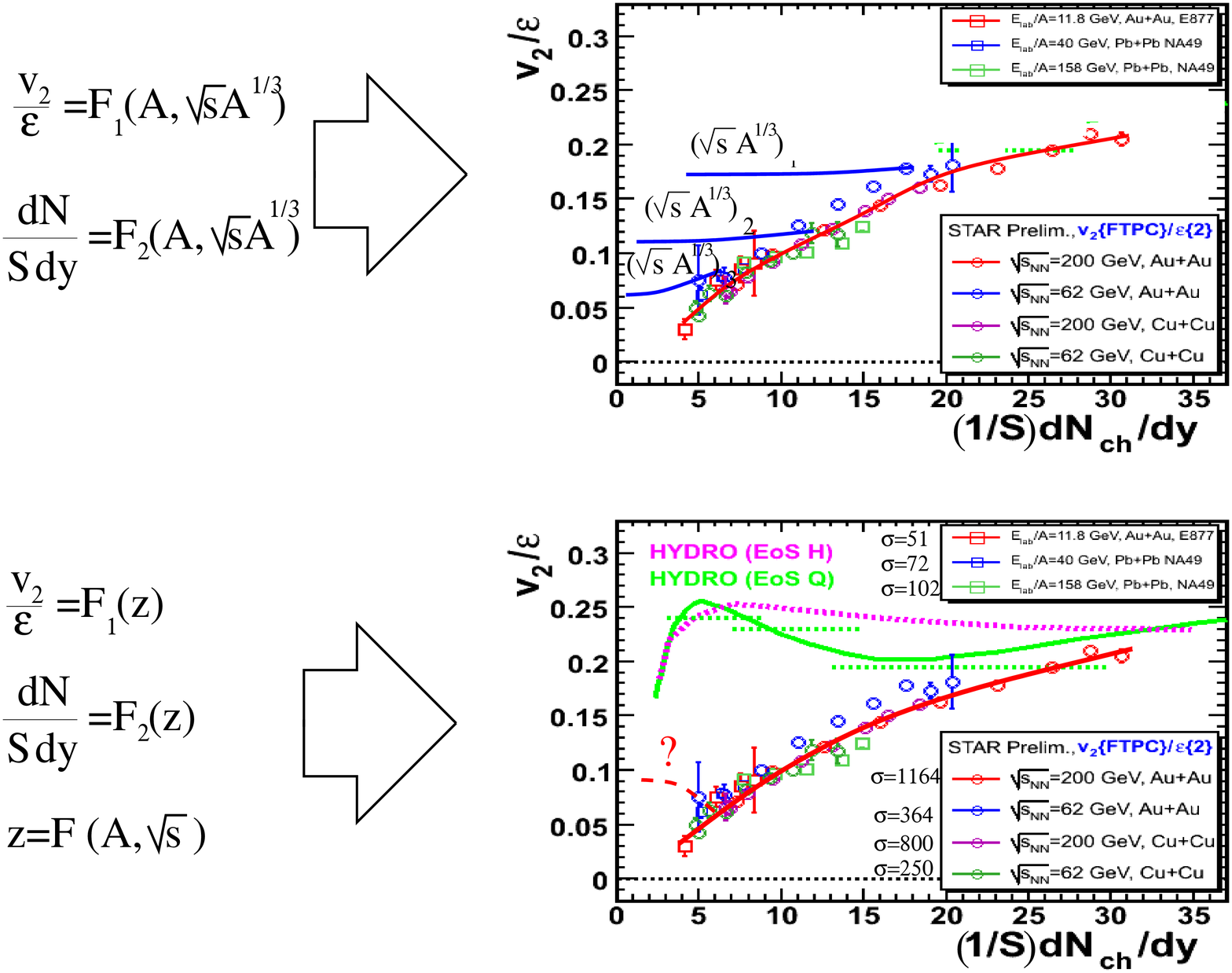}
\caption{(color online)\label{twocurves}An illustration of the mathematical
  implications of universal scaling.  A possible breaking of the
  scaling is shown in the bottom panel}
\end{center}
\end{figure*}

The point made in the previous section is actually straight-forward to understand:
  As illustrated in Fig. \ref{twocurves}
a universal scaling means that the two quantities that scale are
functions of a \textit{ common} variable (that can be, in
general, a function of still further variables).

Thus, the systems from AGS to RHIC appear to be controlled by a
common scale, related to the total multiplicity, which varies smoothly
and drives both $v_2/\epsilon$ and $\frac{1}{S}\frac{dN}{dy}$.
This conclusion is a strong indication that \textit{microscopic}
properties of the system (equation of state and mean free path) are
unchanged, up to a shift related to this scale, in the experimentally accessed energy range.  It also raises
the question of the exact nature of the variable that drives the scaling.

%data implies that initial $l_{mfp}$ and $c_s$ \textit{ do not}
%significantly change over the given range of energy.
%\begin{table}
%\caption{(color online)Energies and system sizes used in heavy ion collisions}
%\begin{tabular}{cccc}
%N & $\sqrt{s}$ (GeV) & A & $\sigma=\sqrt{s} A^{1/3}$ (GeV)  \\ \hline
%\multicolumn{4}{c}{RHIC}\\ \hline
%Cu & 62.5  & 64 &250 \\
%Cu & 200   &64  &800 \\
%Au & 62.5  &197 &363.67 \\
%Au & 130   &197 &756.42 \\
%Au & 200   &197 &1163.73 \\ \hline
%\multicolumn{4}{c}{AGS}\\ \hline
%Au & 4.24  &197 &24.67 \\
%\multicolumn{4}{c}{SPS}\\ \hline
%Pb & 6.26  &207 &37.03 \\
%Pb & 7.21  &207 &42.65 \\
%Pb & 8.76  &207 &51.82 \\
%Pb & 12.32 &207 &72.88 \\
%Pb & 17.27 &207 &102.16 \\
%S  & 19.4 &32 & 61.59 \\ \hline
%\multicolumn{4}{c}{LHC}\\ \hline
%Pb & 2000  &207 & 11830.96\\ \hline
%\end{tabular}
%\end{table}
%It is more natural to suppose that the observed scaling is
%due to a fundamental similarity in all systems under consideration: 
%Intensive properties of these systems are the same, except for one
%overall parameter connected to the total multiplicity.

%Two basic scaling parameters candidates can be considered: One can
%postulate that the system is a 2-fluid model, and what is behind the
%universal scaling is the relative strength of the fluids.
%Alternatively, the scaling can reflect an intensive property of the
%system plus a global size.

%\subsection{QGP fraction}

It has been suggested, in \cite{gyul,shuryak}, that the
system, even at it's initial stage, is not entirely in the nearly
inviscid QGP phase, but a fraction of it is in a highly viscous
hadron gas.  In this case, Eq. \ref{eqv2} and \ref{eqn}
should be updated with new parameters $c_s \Rightarrow c_s^{QGP,HG}$, $l_{mfp} \Rightarrow
l_{mfp}^{QGP,HG}$ and an additional parameter
\begin{equation}
\alpha \sim\frac{ \left( dN/dy \right)_{QGP}}{\left( dN/dy \right)_{total}}
\end{equation}
should be added.

Could it be that $\alpha$ is what moves from zero to unity in the
curve of Fig. \ref{uni}?    If we assume a Glauber model and a Saxon-Wood
distribution for initial density, and a critical ``transition'' energy density
$\rho_c$ (independent of energy and system size), $\alpha$ becomes
\begin{eqnarray}
\label{alpha}
\alpha =\frac{ \int \rho(x,y,z,A,\sqrt{s}) \Theta (\rho(x,y,z,A,\sqrt{s}) -
\rho_c) d^3 x}{\int \rho(x,y,z,A,\sqrt{s}) d^3 x}
\end{eqnarray}
\begin{eqnarray}
\label{rho}
 \rho \sim \frac{\sqrt{s} A^{1/3}}{\ave{z}}  
 \left[
  T_A(\sqrt{(x+b)^2+y^2})+T_B(\sqrt{(x-b)^2+y^2}) \right]
\end{eqnarray}
and $T_{A,B}(r)$ is the usual transverse participant density of the
target nuclei.
\begin{equation}
T(x,y) = \int dz \rho_N (\sqrt{x^2+y^2+z^2} )
\end{equation} 
Such scaling is very non-trivial to model exactly, but it seems to
obey an approximate error function dependence in $\sigma$
\begin{equation}
\alpha \sim \frac{1}{2} \left( 1+ \mathrm{erf} \left(\frac{ \sigma
  f(b) - \rho_c }{\Delta_\sigma }\right)\right)
\end{equation}
it is difficult to see how such a scaling can be compatible with the
observed universal curve.  In particular, there will be limiting
values in $\sqrt{s}$ at which $\alpha$ should go at 0 and 1
independently of $A$ and $b$.   above/below these scales,
the system reduces to a one-component system and branches should start
appearing.

Similarly, at low energies there will be a regime where $\alpha=0$
below a certain impact parameter, and $\alpha>0$ above this impact
parameter.   In the first class of centralities,  the large mean free
path in the ``corona'' would introduce a dependence on $A$
that does not conform to the universal curve.   While RHIC experiments
have isolated several low centrality bins where $v_2$ is significantly
below ideal hydro predictions (the corona dominates), such a critical
centrality has so far not been observed.  (although this can be
explored further with collisions of very heavy nuclei,such as
  Uranium \cite{heinz_uranium,fai}).

We conclude that, if the scaling parameter is indeed the QGP fraction
of the system, the energies explored so far are far from the regime
  where this fraction is either close to unity or to zero.
While this is possible, given the large variation in energy within the
systems explored so to date, it would be somewhat surprising. 

%%\subsection{Knudson number}
%It has been suggested \cite{knudsen} that the only relevant
%variable within $v_2$ calculations is the Knudsen number, defined
%simply as the number of collisions between the degrees of freedom of
%the system
%\begin{equation}
%K \sim \frac{R}{l_{mfp}} 
%\end{equation}
%As we have argued, the observed scaling is very difficoult to
%reconcile with the threshold needed for a phase transition:
%While $R$ depends trivially on the system size, $l_{mfp}$ depends on
%the initial energy density.
%
%Hence, if an above-threshold and a below threshold system were
%observed, they could have two very different Knudson numbers but the same
%$\frac{1}{S} \frac{dN}{dy}$.  This, together with the different
%equation of state, would in general lead to different $v_2/\epsilon$.
%We can therefore conclude that either the Knudson number is not a
%relevant parameter, or the systems produced are all within the same
%phase within the scaling region considered. 
%\subsection{initial conditions}

An alternative ansatz to reproduce the observed scaling is by
postulating a universality in initial conditions, up to a scale
parameter, that also smoothly controls the initial temperature (and
hence $c_s$ and $l_{mfp}$). 
The basic constraint that the universal scaling imposes on a set of
uniform initial conditions is the requirement that a single
dimensionful scale $\ave{\tau}$ exists, and varying \textit{either} the energy or
the system size only shifts the system up and down the scale $\ave{\tau}$.
Both $\ave{z} n(T,\mu_B)$ and $v_2$ are then functions of only $\ave{\tau}$
as well as constants independent of energy and system size.
A natural interpretation for
$\ave{\tau}$ is the system's lifetime in the co-moving frame.    
In units of the mean free path, this corresponds to the inverse of the Knudsen number
\cite{knudsen}, the number of collisions between the system's degrees
of freedom.

Because of the leading dependence of
$\frac{1}{S}\frac{dN}{dy}$ on $\ave{z} n(T,\mu_B)$ (Eq. \ref{eqn}), it follows that
$\ave{\tau}$ can only scale with $A,\sqrt{s},\sigma$ in the same way as
$\ave{z} n(T,\mu_B)$.   
\begin{equation}
\ave{\tau} = F^{-1} \left(\ave{z} n(T,\mu_B) \right)
\end{equation}
where $F(\ave{\tau})$ is the same for all energies. A different dependence would lead to two
different $\ave{\tau}$s corresponding to the same $\ave{z} n(T,\mu_B)$,
which would again break the observed scaling.
 
Eq. \ref{eqv2} and \ref{eqn} then simplify to
\begin{eqnarray}
v_2 \sim \epsilon \left( a_{100} + a_{101} c_s + a_{102} c_s^2 + ... \right) + \nonumber \\
      \frac{l_{mfp}}{\ave{\tau}}\epsilon  \left(  a_{110}  + a_{111}
       c_s +... \right) + \nonumber \\
      \left( \frac{l_{mfp}}{\ave{\tau}} \right)^2 \epsilon \left( a_{120} + a_{121}
      c_s +...  \right)+...
\label{eqv2reduced}
\end{eqnarray}
\begin{eqnarray}
\frac{1}{S} \frac{dN}{dy} \sim F(\ave{\tau}) \left( 1 +  b_{010}
 \frac{ l_{mfp}}{\ave{\tau}} + b_{011} \frac{c_s l_{mfp}}{\ave{\tau}} + \right. 
 \nonumber \\
 \left. b_{2 00}  \left( \frac{l_{mfp}}{\ave{\tau}} \right)^2+ ... \right)
\label{eqnreduced}
\end{eqnarray}
where $l_{mfp}$,$b$,$a$ depend only on $\ave{\tau}$, presumably via a
simple scaling between $\ave{\tau}$ and initial temperature.
This scaling will most probably be monotonic
\begin{equation}
\ave{\tau} = F \left( \frac{T_{freeze-out}}{T_{initial}} \right) \sim \sigma^n
\end{equation}
More complicated scalings, with minima and sharp transitions, will in
general lead to a violation of the universal scaling, since events
with similar final multiplicities could in this case have different
$v_2/\epsilon$.

To fully appreciate these constraints, it must be remembered that the
lifetime strongly depends on the system's longitudinal initial
conditions, and in particular Landau and Bjorken type initial conditions, with
the same equation of state and transport coefficients, will lead to
very different $\ave{\tau}$s \cite{obscurica}.

Thus, the scaling of $v_2$ rules out a transition of the system from
the Landau to the Bjorken limit such as the one seen in the top
panel of Fig. \ref{sqrts}, in the considered range of energies and
system sizes:  Such a transition would mean that two events with the same
$\frac{1}{S}\frac{dN}{dy}$, one high-energy non-central,
the other low-energy central, would correspond to two different
lifetimes (the first close to the Bjorken limit,the second to the
Landau limit), and hence, in general, to two $v_2/\epsilon$.  

The monotonic increase of $v_2/\epsilon$ further constrains either the
initial conditions to be far away from the Bjorken limit, or the mean
free path to be non-negligible, at all energies :  
As initially inferred in \cite{v2orig}, and explicitly shown in \cite{shuryak,heinz,huovinen} $v_2$ is a \textit{self-quenching}
signature, which saturates after a finite time $\tau_{v2}$, with $\tau_{v2} \ll \ave{\tau}$ in the Bjorken limit
\cite{shuryak,heinz,huovinen}.
The $v_2$ scaling than implies that the system never reaches
$\tau_{v2}$.  If it did, systems with different
$\ave{\tau}_{1,2}>\tau_{v2}$ but the same $\epsilon$
would have the same $v_2/\epsilon$.   It is clear that such systems
would, in general, have very different $\frac{1}{S}\frac{dN}{dy}$,
breaking the scaling.

The universal scaling of $v_2$ in pseudo-rapidity space observed in \cite{scalingptphobos} adds a further layer of constraints to Eq. \ref{eqv2reduced}.It appears that the $\sigma$ is also a uniquely determined universal function of the relative position of the volume element in rapidity space.  This rules out a ``Landau''/fire-strak model where the system is closely localized in rapidity.
Instead, the fireball evolves in a way that is both local in rapidity, and strongly rapidity-dependent.
Perhaps the BGK initial condition \cite{bgk}, could provide such an ansatz, although it would imply that such a geometry holds, to a good approximation, up to AGS energies.

The picture suggested by the ``universal'' scaling of $v_2$ is then
considerably different from the ``RHIC reaches the perfect fluid''
scenario:
The ``fluid'' produced in heavy ion collisions, at all energies and
system sizes where the scaling holds, should have a comparable $l_{mfp}$,
equation of state ($c_s$) and longitudinal structure of the initial
condition.
All that varies is lifetime $\ave{\tau}$, which is uniquely
determined by the initial density multiplied by longitudinal size,
$\ave{z} n(T,\mu_B)$.    This universal structure is robust inasmuch
the scaling experimentally observed.

   The alternative is that ``we all have got it wrong'' and the picture quantitatively analyzed in \cite{voloshin1}, of a very weakly interacting system, is the appropriate one for describing heavy ion collisions, from AGS to RHIC.
But, aside from the difficulty in modeling such a large $v_2$ in this picture, the conclusions in the previous paragraph would actually {\em not} be changed:
For the scaling shown in \cite{voloshin1} to hold, it is necessary for $\ave{v_{ij} \sigma_{ij}}$, where $v_{ij}$ is the relative speed and $\sigma_{ij}$ the cross-sectional area of the system's microscopic degrees of freedom, not to change significantly with energy and system size.  This is equivalent to requiring the microscopic properties of the system, such as viscosity and equation of state, to remain the same.

Experimentally it will be very interesting to see at what point, in
\textit{ low} energy collisions, is the universal scaling observed here
broken.   This point could well be the critical $\sigma$ that
produces a deconfined system.     

Perhaps a greater energy and system size
exploration around the region of the so called
``kink'',''horn'',''step'' \cite{horn} anomaly can yield discoveries;  As
seen in Fig. \ref{twocurves} (lower panel,
$\frac{1}{S}\frac{dN}{dy}\sim 6 fm^{-2}$), there might be a hint of splitting in the
scaling curve;  While the error bars abundantly drown out any firm evidence
at this point, the approximate coincidence with the features highlighted in
\cite{horn}, as well as the breaking of $HBT$ radii scaling with 
multiplicity and $\sqrt{s}$ \cite{lisahbt} are suggestive.
Likewise, it will be interesting to see if peripheral LHC collisions,
where the initial temperature should lie solidly in the asymptotic
freedom regime, can be related to central top energy RHIC collisions.

The observation that the scaling could apply not just with integrated $p_T$ but within 
$p_T$ bins (\cite{scalingphenix,scalingptphobos}) also deserves further exploration.
One could object that, since $\ave{p_T}$ is energy dependent, a meaningful 
comparison across different energy regimes can not be obtained.  However, in the 
hydrodynamic picture $\ave{p_T}$ is also strongly system-size dependent because of the 
growth of transverse flow, yet \cite{scalingphenix,scalingptphobos} finds a $p_T$ 
specific scaling 
to hold when Cu-Cu and Au-Au are compared.   We await further, energy dependent results in this direction, but remark that the breaking of this scaling could signal the energy scale at which equilibration stops applying.
 
%\textit{if} the mean free path of the system is non-negligible, than $R_0$ is a constant independent of the energy.   This is a very
%surprising result, since it would indicate that the \textit{initial
%  state} dynamics at the considered energies does not exhibit
%\textit{qualitative} differences, but only a quantitative shift
%related, over all energies, by a common scale.
%Nevertheless, it
%is important to realize that Eq. \ref{initial} is, indicated, in
%a model-independent way, by experimental data:  \textit{if}

%%%%%%%%%%%%%%%%%%%%%%%%%%%%%%
%%%%%%%%%%%%%%%%%%%%%%%%%%%%%%%

\section{conclusion}
In conclusion, we have shown that the scaling of $v_2/\epsilon$ with
$\frac{1}{S}\frac{dN}{dy}$ places tight constraints on the
hydrodynamic initial conditions in heavy ion collisions.  
It imposes an energy-independent relationship between initial energy
density and longitudinal size, and makes it likely
that longitudinal structure and microscopic parameters, such as the initial temperature, equation
of state and viscosity are comparable in the considered range of
energies and system sizes.  We have suggested that looking for when the given scaling breaks might yield information about the
critical energy and system size at which we can speak of a deconfined
collective phase.

The author would like to thank the Von Humboldt foundation and Frankfurt
university for the support provided, the institute for
Nuclear Physics in Krakow for hospitality in the days most of this
paper was written, and Miklos Gyulassy, Sangyong Jeon, Sergei Voloshin 
Wojciech Florkowski, Wojciech Broniowski and Piotr Bozek  for useful discussions.

\end{document}